\documentclass[preprintnumbers,nofootinbib,superscriptaddress,10pt]{revtex4}
\usepackage[dvipdfmx]{graphicx} 
\usepackage{amsmath,amssymb,amsfonts, bm} 

\def\a{\alpha} \def\b{\beta} \def\g{\gamma}  \def\d{\delta} \def\D{\Delta}    \def\th{\theta}                     

  \def\nn{\nonumber}

\usepackage{color}

\begin{document}

\title{\large Rephasing invariant formulae for CP phases in general parameterizations of flavor mixing matrix 
and  exact sum rules with unitarity triangles }

\preprint{STUPP-25-285}

\author{Masaki J. S. Yang}
\email{mjsyang@mail.saitama-u.ac.jp}
\affiliation{Department of Physics, Saitama University, 
Shimo-okubo, Sakura-ku, Saitama, 338-8570, Japan}
\affiliation{Department of Physics, Graduate School of Engineering Science,
Yokohama National University, Yokohama, 240-8501, Japan}



\begin{abstract} 

In this letter, we present rephasing invariant formulae
$\delta^{(\alpha i)} = \arg [ { V_{\alpha 1} V_{\alpha 2} V_{\alpha 3} V_{1i} V_{2i} V_{3i} / V_{\alpha i }^{3} \det V } ] $
 for CP phases $\delta^{(\alpha i)}$ associated 
with nine Euler-angle-like parameterizations of a flavor mixing matrix.  
Here, $\alpha$ and $i$ denote the row and column carrying the trivial phases in a given parameterization.
Furthermore, we show that the phases $\delta^{(\alpha i)}$ and the nine angles $\Phi_{\alpha i}$ of unitarity triangles satisfy compact sum rules 
$ \delta^{(\alpha, i+2)} - \delta^{(\alpha, i+1)} = \Phi_{\alpha+1, i} - \Phi_{\alpha+2, i}$
and
$ \delta^{(\alpha+1, i)} - \delta^{(\alpha+2, i)} = \Phi_{\alpha, i+2} - \Phi_{\alpha, i+1}$
where all indices are taken cyclically modulo three. 
These relations are natural generalizations of the previous result 
$\delta_{\mathrm{PDG}}+\delta_{\mathrm{KM}}=\pi-\alpha+\gamma.$

\end{abstract} 

\maketitle

\section{Introduction}

The flavor mixing of quarks and leptons has been one of the essential pieces 
for the development of particle physics, and it is also an intrinsically interesting subject of mathematical investigation. 
Among various representations of the mixing matrix \cite{Kobayashi:1973fv, Chau:1984fp, Hall:1993ni, Fritzsch:1995nx, Kaur:2023ypg, Yang:2020qsa, Yang:2020goc, Yang:2024ulq, Yang:2025yst}, 
Euler-angle-like parameterizations in terms of three rotation matrices and a distinct CP phase admit  nine possible forms \cite{Rasin:1997pn}.  
However, at first glance, these nine phases seem to take chaotic values, showing no apparent pattern.

On the other hand, unitarity conditions of the mixing matrix give rise to six unitarity triangles  \cite{Wu:1994di, Xing:2009eg, Harrison:2009bz, He:2013rba, He:2016dco, Xing:2019tsn, Harrison:2025rkp} and 
a lot of papers have also been studied
relations between the CP phases and the triangles \cite{Koide:2004gj, Koide:2008yu, Frampton:2010ii, Dueck:2010fa, Frampton:2010uq, Li:2010ae, Qin:2011bq, Zhou:2011xm, Qin:2011ub, Qin:2010hn, Zhang:2012ys, Li:2012zxa, Zhang:2012bk}. 
These nine angles also seem to have independent values.  

More recently,  
rephasing invariant formulae for the CP phases \cite{Yang:2025hex,Yang:2025law,Yang:2025ftl}
produce a simple sum rule relating the CP phases and angles 
$\delta_{\rm PDG}+\delta_{\rm KM}=\pi-\alpha+\gamma$. 
This led to the discovery of an underlying order between the random CP phases and angles. 

In this letter, to organize relations connecting the nine CP phases and the nine angles of unitarity triangles, 
we first derive rephasing invariant formulae for each CP phase in terms of elements of the mixing matrix $V$ and $\det V$. 
Next, we present exact sum rules between the phases and angles, demonstrating that the previous sum rule is a special case of these relations.

\section{Rephasing invariant formulae for CP phases in general parameterizations of flavor mixing matrix}

In this section, following our previous works \cite{Yang:2025hex, Yang:2025law, Yang:2025ftl}, 
we derive rephasing invariant formulae for CP phases in Euler-angle-like parameterizations of the flavor mixing matrix $V$.
For this purpose, we adopt the nine parameterizations proposed by Fritzsch and Xing \cite{Fritzsch:1997st}.

We begin by defining the $2\times 2$ rotation matrix $R_{ij}$ as follows, 
\begin{align}
R_{12}(\theta)  =
\begin{pmatrix}
c_{\theta} & s_{\theta} & 0 \\
- s_{\theta} & c_{\theta} & 0 \\
0 & 0 & 1 \\
\end{pmatrix} , ~~~
R_{23}(\sigma)  = 
\begin{pmatrix}
1 & 0 & 0 \cr
0 & c_{\sigma} & s_{\sigma} \\
0 & - s_{\sigma} & c_{\sigma} \\
\end{pmatrix} , ~~ 
R_{31} (\tau)  =
\begin{pmatrix}
c_{\tau} & 0 & s_{\tau} \\
0  & 1 & 0 \\
- s_{\tau} & 0 & c_{\tau} \\
\end{pmatrix} , 
\label{Rij}
\end{align}
where $s_{\th} = \sin \th , c_{\th} = \cos \th. $ 
Furthermore, complex rotation matrices $R_{12}(\theta, \delta)$, $R_{23}(\sigma, \delta)$, and $R_{31}(\tau, \delta)$ are defined by replacing $1 \rightarrow e^{-i\delta}$ in Eq.~(\ref{Rij}).
With these matrices, the flavor mixing matrix $V$ admits nine different parameterizations \cite{Fritzsch:1997st}
\begin{align}
P1: ~ V^{(33)} &= R_{12}(\theta) R_{23}(\sigma, \d^{(33)}) R^{-1}_{12}(\theta') \, , ~~
P2: ~ V^{(11)} = R_{23}(\sigma) R_{12}(\theta, \d^{(11)}) R^{-1}_{23}(\sigma') \, , \nn \\
P3: ~ V^{(13)} &= R_{23}(\sigma) R_{31}(\tau, \d^{(13)})  R_{12}(\theta) \, , ~~~ 
P4: ~ V^{(31)} = R_{12}(\theta) R_{31}(\tau, \d^{(31)})  R^{-1}_{23}(\sigma) \, ,  \nn \\
P5: ~ V^{(22)} &= R_{31}(\tau) R_{12}(\theta, \d^{(22)}) R^{-1}_{31}(\tau') \, , ~~ 
P6: ~ V^{(32)} = R_{12}(\theta) R_{23}(\sigma, \d^{(32)}) R_{31}(\tau) \, , \nn \\
P7: ~ V^{(12)} &= R_{23}(\sigma) R_{12}(\theta, \d^{(12)}) R^{-1}_{31}(\tau) \, ,  ~~ 
P8: ~ V^{(21)} = R_{31}(\tau) R_{12}(\theta, \d^{(21)}) R_{23}(\sigma) \, ,  \nn \\
P9: ~ V^{(23)} &= R_{31}(\tau)  R_{23}(\sigma, \d^{(23)}) R^{-1}_{12}(\theta) \, . 
\end{align}
Here, in each of these parameterizations, a row $\alpha$ and a column $i$ have trivial phases, $V_{\alpha j}, V_{\beta i} \in \mathbb{R}$. 
The notation $V^{(\alpha i)}$ and $\delta^{(\alpha i)}$ distinguish the corresponding matrix and CP phase.  
More explicitly, the conditions are
\begin{align}
\arg V_{\a 1}^{(\a i)} = \arg V_{\a 2}^{(\a i)} = \arg V_{\a 3}^{(\a i)} = \arg V_{1i}^{(\a i)} = \arg V_{2 i }^{(\a i )} = \arg V_{3 i}^{(\a i )} =  0 ~ {\rm or} ~ \pi \, . 
\end{align}
Since there is a duplication for the element $V_{\alpha i}^{(\alpha i)}$, there are effectively five independent conditions, and the phase $\pi$ appears either zero or two times. 
As the final condition, the argument of the determinant is
\begin{align}
\arg  \det V^{(\a i )} = - \d^{(\a i )} \, .
\end{align}
In other words, the phase structures of these parameterizations $V^{(\alpha i)}$ are specified by these six conditions.

Let us consider transforming the mixing matrix $V$ in an arbitrary basis into one of these parameterizations $V^{(\alpha i)}$.
Suppose that general rephasing transformations remove unphysical phases as
\begin{align}
V^{(\alpha i)} = \Psi_{L}^{\dagger} V \Psi_{R},
\end{align}
where $(\Psi_{L})_{\alpha\beta} = e^{i \gamma_{L \alpha}} \delta_{\alpha \beta}$ and $(\Psi_{R})_{ij} = e^{i \gamma_{R i}} \delta_{ij}$ are diagonal phase matrices.
The inverse transformation from $V^{(\alpha i)} $ to the original matrix $V$ is given by
\begin{align}
V = \Psi_{L} V^{(\a i )}  \Psi_{R} ^{\dagger} \, , ~~
V_{\b j} = e^{i \g_{L \b}} V^{(\a i)}_{\b j} e^{- i \g_{R j}} \, . 
\end{align}
By taking arguments of the determinants,
a relation $\arg \det V = - \delta^{(\alpha i)} + \sum_{\beta, j} (\gamma_{L \beta} - \gamma_{R j})$ is 
obtained and the CP phase $\delta^{(\alpha i)}$ is found to be
\begin{align}
\d^{(\a i )} & =  (\g_{L1} + \g_{L2} + \g_{L3} - \g_{R1} - \g_{R2} - \g_{R3}) - \arg  \det V \nn \\
& = 
\arg [V_{\a 1} V_{\a 2} V_{\a 3} V_{1i} V_{2i} V_{3i} / V_{\a i}^{3}] - \arg  \det V
= \arg \left[ { V_{\a 1} V_{\a 2} V_{\a 3} V_{1i} V_{2i} V_{3i}  \over V_{\a i }^{3} \det V } \right] . 
\end{align}
In practice, since $V_{\alpha i}$ appears twice in the numerator, 
cancellations occur between the numerator and denominator as
\begin{align}
\d^{(\a i )} = \arg \left[ { V_{\a 1} V_{\a 2} V_{\a 3} V_{1i} V_{2i} V_{3i}  \over V_{\a i }^{3} \det V } \right] 
= \arg \left[  {  ( \prod_{j \neq i} V_{\a j} ) \, ( \prod_{\b \neq \a}  V_{\b i} )  \over V_{\a i } \det V } \right] . 
\end{align}
Since six phases are required to eliminate unphysical phases  $\gamma_{L \alpha}$ and $\gamma_{R i}$, 
any other combinations have nontrivial phases of $V^{(\a i )}_{\b j }$ and are not solved for $\d^{(\a i)}$ explicitly.  
This derivation remains valid even in the presence of Majorana phases for leptons
because the resulting CP phases are manifestly rephasing invariant. 

Explicitly, the formulae are expressed as 
\begin{align}
\d^{(11)} & = \arg \left[ { V_{12} V_{13}  V_{21} V_{31}  \over V_{11} \det V } \right]  , ~~ 
\d^{(12)} = \arg \left[ { V_{11} V_{13}  V_{22} V_{32}  \over V_{12} \det V} \right] , ~~
\d^{(13)} = \arg \left[ { V_{11} V_{12}  V_{23} V_{33}  \over V_{13} \det V} \right]  , \nn \\
\d^{(21)} & = \arg \left[ { V_{22} V_{23} V_{11}  V_{31}  \over V_{21} \det V} \right]  , ~~ 
\d^{(22)} = \arg \left[ {V_{21} V_{23}V_{12}  V_{32}   \over V_{22} \det V} \right]  , ~~
\d^{(23)} = \arg \left[ { V_{21} V_{22} V_{13} V_{33}  \over V_{23} \det V} \right]  , \nn \\
\d^{(31)} & = \arg \left[ { V_{32} V_{33} V_{11} V_{21}   \over V_{31}\det V} \right]  , ~~ 
\d^{(32)} = \arg \left[ { V_{31} V_{33}  V_{12} V_{22}  \over V_{32} \det V} \right]  , ~~
\d^{(33)} = \arg \left[ { V_{31} V_{32} V_{13} V_{23}  \over V_{33} \det V} \right] . 
\end{align}
They provide a necessary and sufficient set of
 ``irreducible fifth-order'' invariants which cannot be decomposed into second- and third-orders,  
because only nine such invariants are constructed using $\det V$. 

From the latest UTfit results of the CKM matrix parameters  \cite{UTfit:2022hsi}
\begin{align}
\sin \th_{12} &= 0.22519 \pm 0.00083 \, ,  ~~~ \sin \th_{23} = 0.04200 \pm 0.00047 \, ,  \nn \\
\sin \th_{13} &= 0.003714 \pm 0.000092 \, ,  ~~~ \d = 1.137 \pm 0.022  = 65.15^{\circ} \pm1.3^{\circ}  \, ,
\end{align}
the numerical evaluation at the best-fit values yields
\begin{align}
\begin{pmatrix}
\d^{(11)} & \d^{(12)} & \d^{(13)} \\
\d^{(21)} & \d^{(22)} & \d^{(23)} \\
\d^{(31)} & \d^{(32)} & \d^{(33)} 
\end{pmatrix}
=
\begin{pmatrix}
 92.44^{\circ} & 115.91^{\circ} & 65.15^{\circ} \\
 157.51^{\circ} & 1.089^{\circ} & 114.89^{\circ} \\
 23.54^{\circ} & 156.49^{\circ} & 93.45^{\circ} \\
\end{pmatrix} .
\end{align}
In particular, the standard PDG phase $\delta_{\rm PDG}$
 and the original Kobayashi--Maskawa phase $\delta_{\rm KM}$ are given by
\begin{align}
\d_{\rm PDG} = \d^{(13)} \, , ~~~ \d_{\rm KM} = \pi - \d^{(11)} \, . 
\end{align}
Interestingly, a sum over any row or column of the phases $\delta^{(ij)}$ 
 yields the same value of a ninth-order invariant 
\begin{align}
\sum_{i} \d^{(ij)} = \sum_{j} \d^{(ij)} = 
 \arg \left[ { \prod_{\a, i} V_{\a i} \over \det V^{3} } \right] = 273.49^{\circ} \, . 
 \label{sumofphases}
\end{align}

For a cross check, we numerically compute $\cos \delta^{(ij)}$,
\begin{align}
\begin{pmatrix}
\cos \d^{(11)} &\cos \d^{(12)} & \cos \d^{(13)} \\
\cos \d^{(21)} &\cos \d^{(22)} & \cos \d^{(23)} \\
\cos \d^{(31)} &\cos \d^{(32)} & \cos \d^{(33)} 
\end{pmatrix}
=
\begin{pmatrix}
 -0.0425 & -0.437 & 0.420 \\
 -0.924 & 0.9998 & -0.421 \\
 0.917 & -0.917 & -0.0603 \\
\end{pmatrix} .
\end{align}
These values agree with values obtained from moduli of matrix elements, 
\begin{align}
\cos \d^{(11)} &= \frac{\left(1-| V_{11}| ^2\right)^2 | V_{22}| ^2 -| V_{11} V_{12} V_{21}| ^2-| V_{13} V_{31}| ^2}{2 | V_{11} V_{12} V_{13} V_{21} V_{31}| } = -0.0425 \, , \nn\\
\cos \d^{(12)} &= \frac{\left(1-| V_{12}| ^2\right)^2 | V_{23}| ^2 -| V_{12} V_{13} V_{22}| ^2-| V_{11} V_{32}| ^2}{2 | V_{11} V_{12} V_{13} V_{22} V_{32}| } = -0.437 \, , \nn \\
\cos \d^{(13)} &= -\frac{\left(1-| V_{13}| ^2\right)^2 | V_{22}| ^2-| V_{12} V_{23} V_{13}| ^2-| V_{11} V_{33}| ^2}{2 | V_{11} V_{12} V_{13} V_{23} V_{33}| } = 0.420 \, , \nn \\
\cos \d^{(21)} &= -\frac{ \left(1-| V_{21}| ^2\right)^2| V_{12}| ^2 -| V_{11} V_{21} V_{22}| ^2-| V_{23} V_{31}| ^2}{2 | V_{21} V_{22} V_{23} V_{11} V_{31}| } = -0.924 \, , \nn \\
\cos \d^{(22)} & = \frac{\left(1-| V_{22}| ^2\right)^2 | V_{33}| ^2 -| V_{22} V_{23} V_{32}| ^2 -| V_{12} V_{21}| ^2}{2 | V_{21} V_{22} V_{23} V_{12} V_{32}| } =  0.9998 \, , \nn \\
\cos \d^{(23)} &= -\frac{\left(1-| V_{23}| ^2\right)^2 | V_{32}| ^2 -| V_{22} V_{23} V_{33}| ^2 -| V_{13} V_{21}| ^2}{2 | V_{21} V_{22} V_{23} V_{13} V_{33}| } = -0.421 \, , \nn \\
\cos \d^{(31)} &= -\frac{ \left(1-| V_{31}| ^2\right)^2 | V_{22}| ^2 - | V_{21} V_{32} V_{31}| ^2-| V_{11} V_{33}| ^2}{2 | V_{11} V_{21} V_{31} V_{32} V_{33}| } = 0.917 \, , \nn \\
\cos \d^{(32)} &= -\frac{ \left(1-| V_{32}| ^2\right)^2 | V_{23}| ^2 -| V_{22} V_{32} V_{33}| ^2 -| V_{12} V_{31}| ^2}{2 | V_{31} V_{32} V_{33} V_{12} V_{22}| } = -0.917 \, , \nn \\
\cos \d^{(33)} &= \frac{ \left(1-| V_{33}| ^2\right)^2 | V_{11}| ^2 -| V_{13} V_{31} V_{33}| ^2 -| V_{23} V_{32}| ^2}{2 | V_{13} V_{31} V_{33} V_{23} V_{32}| } = -0.0603 \, .  
\end{align}
Since $\delta^{(\a i )}$ also carries the sign of $\sin \delta^{(\a i)}$, it contains more information than such an evaluation based on absolute values of matrix elements. 

\section{Exact sum rules between CP phases in general parameterizations and  unitarity triangles}

It is noteworthy to reconsider relations between the results and angles of unitarity triangles.  
The three angles of the unitarity triangle are rephasing invariants,
\begin{align}
\a = \arg \left [ - { V_{td}^{} V_{tb}^{*} \over V_{ud}^{} V_{ub}^{*} } \right ] = 92.40^{\circ}, ~~
\b = \arg \left [ - { V_{cd}^{} V_{cb}^{*} \over V_{td}^{} V_{tb}^{*} }  \right ] = 22.49^{\circ}, ~~
\g = \arg \left [ - { V_{ud}^{} V_{ub}^{*} \over V_{cd}^{} V_{cb}^{*} }  \right ] = 65.11 ^{\circ} . 
\end{align}
A general phase matrix $\Phi$ of these angles is defined in \cite{Harrison:2009bz}. 
In particular, the elements in the second column of $\Phi$ correspond to $\beta$, $\alpha$, and $\gamma$, respectively.
\begin{align}
\Phi 
\equiv 
\begin{pmatrix}
\arg  (-\Pi_{ud}^*) & \arg (-\Pi_{us}^*) & \arg  (-\Pi_{ub}^*) \\
\arg  (-\Pi_{cd}^*) & \arg  (-\Pi_{cs}^*) & \arg  (-\Pi_{cb}^*) \\
\arg  (-\Pi_{td}^*) & \arg  (-\Pi_{ts}^*) & \arg  (-\Pi_{tb}^*) \\
\end{pmatrix}
=
\begin{pmatrix}
 1.054^{\circ} & 22.49^{\circ} & 156.46^{\circ} \\
 64.09^{\circ} & 92.40^{\circ} & 23.50^{\circ} \\
 114.85^{\circ} & 65.11^{\circ} & 0.0370^{\circ} \\
 \end{pmatrix} .
\end{align}
Here, $\Pi_{\alpha i} \equiv V_{\beta j} V_{\beta k}^* V_{\gamma k} V_{\gamma j}^* $
and the indices are defined cyclically.
For better readability, all flavor indices are rewritten by using numbers as
\begin{align}
\Pi = 
\begin{pmatrix}
  V_{33} V_{32}^* V_{22} V_{23}^*
& V_{31} V_{33}^* V_{23} V_{21}^*
& V_{32} V_{31}^* V_{21} V_{22}^*  \\
   V_{13} V_{12}^* V_{32} V_{33}^*
& V_{11} V_{13}^* V_{33} V_{31}^*
& V_{12} V_{11}^* V_{31} V_{32}^*   \\
   V_{23} V_{22}^* V_{12} V_{13}^*
& V_{21} V_{23}^* V_{13} V_{11}^*
& V_{22} V_{21}^* V_{11} V_{12}^* 
\end{pmatrix} . 
\end{align}
The sum over each row and column of the matrix $\Phi$ equals $180^{\circ}$, 
\begin{align}
\sum_{i} \Phi_{ij} = \sum_{j} \Phi_{ij} = \pi \, .  
\label{sumofangles}
\end{align}

Since solving general sum rules is a challenging task, we employ a transfer matrix approach.  
Defining a phase matrix $\D$ and a transfer matrix $T$, 
\begin{align}
\D \equiv 
\begin{pmatrix}
\d^{(11)} & \d^{(12)} & \d^{(13)}  \\
\d^{(21)} & \d^{(22)} & \d^{(23)}  \\
\d^{(31)} & \d^{(32)} & \d^{(33)} 
\end{pmatrix} , 
~~ 
T \equiv 
\begin{pmatrix}
0 & 0 & 1 \\
1 & 0 & 0 \\
0 & 1 & 0 \\
\end{pmatrix}
, 
\end{align}
one finds the following relations 
\begin{align}
 \D T - \D = 
\begin{pmatrix}
 \arg \frac{V_{11}^2 V_{22} V_{32}}{V_{12}^2 V_{21} V_{31}} &  \arg \frac{V_{12}^2 V_{23} V_{33}}{V_{13}^2 V_{22} V_{32}} &  \arg \frac{V_{13}^2 V_{21} V_{31}}{V_{11}^2 V_{23} V_{33}} \\
  \arg \frac{V_{12} V_{21}^2 V_{32}}{V_{11} V_{22}^2 V_{31}} &  \arg \frac{V_{13} V_{22}^2 V_{33}}{V_{12} V_{23}^2 V_{32}} & \arg \frac{V_{11} V_{23}^2 V_{31}}{V_{13} V_{21}^2 V_{33}} \\
 \arg \frac{V_{12} V_{22} V_{31}^2}{V_{11} V_{21} V_{32}^2} & \arg \frac{V_{13} V_{23} V_{32}^2}{V_{12} V_{22} V_{33}^2} & \arg \frac{V_{11} V_{21} V_{33}^2}{V_{13} V_{23} V_{31}^2} \\
\end{pmatrix}
&  = T^{2}  \Phi T^{2} -T \Phi T^{2} \, . 
\end{align}
Alternatively, by multiplying $T$ from the right and using $T^{2} = T^{T} = T^{-1}$, 
it leads to a more compact notation 
\begin{align}
 \D T^{2} - \D T = T^{2}  \Phi - T \Phi  \, .
\end{align}
This represents the transfer along columns of $\D$ and rows of $\Phi$; 
\begin{align}
 \d^{(\a , i+2)} - \d^{(\a, i+1)} = \Phi_{\a+1, i} - \Phi_{\a+2, i} \, . 
 \label{Tcolumn}
\end{align}
Here, the indices are taken modulo three. 
Note that the transfer $T^{T}$ from the left increases the indices of rows.   
These relations are  numerically verified as 
\begin{align}
&
\begin{pmatrix}
 65.15^{\circ} &  92.44^{\circ} & 115.91^{\circ} \\
114.89^{\circ} &  157.51^{\circ} & 1.089^{\circ}  \\
 93.45^{\circ} &  23.54^{\circ} & 156.49^{\circ}  \\
\end{pmatrix}
-
\begin{pmatrix}
 115.91^{\circ} &  65.15^{\circ} &  92.44^{\circ} \\
 1.089^{\circ} & 114.89^{\circ} &  157.51^{\circ}  \\
 156.49^{\circ} &  93.45^{\circ} &  23.54^{\circ}  \\
\end{pmatrix}
= 
\begin{pmatrix}
 -50.76 & 27.29 & 23.47 \\
 113.80 & 42.62 & -156.42 \\
 -63.04 & -69.91 & 132.95 \\
\end{pmatrix} \nn \\
& = 
\begin{pmatrix}
 64.09^{\circ} & 92.40^{\circ} & 23.50^{\circ} \\
 114.85^{\circ} & 65.11^{\circ} & 0.0370^{\circ} \\
 1.054^{\circ} & 22.49^{\circ} & 156.46^{\circ} \\
 \end{pmatrix} 
-
\begin{pmatrix}
 114.85^{\circ} & 65.11^{\circ} & 0.0370^{\circ} \\
 1.054^{\circ} & 22.49^{\circ} & 156.46^{\circ} \\
  64.09^{\circ} & 92.40^{\circ} & 23.50^{\circ} \\
 \end{pmatrix} .
\end{align}

Similarly, we can find another transfer as
\begin{align}
 T \D - \D = 
\begin{pmatrix}
\arg \frac{V_{11}^2 V_{32} V_{33}}{V_{12} V_{13} V_{31}^2} & \arg \frac{V_{12}^2 V_{31} V_{33}}{V_{11} V_{13} V_{32}^2} &\arg \frac{V_{13}^2 V_{31} V_{32}}{V_{11} V_{12} V_{33}^2} \\
\arg \frac{V_{12} V_{13} V_{21}^2}{V_{11}^2 V_{22} V_{23}} &\arg \frac{V_{11} V_{13} V_{22}^2}{V_{12}^2 V_{21} V_{23}} &\arg \frac{V_{11} V_{12} V_{23}^2}{V_{13}^2 V_{21} V_{22}} \\
\arg \frac{V_{22} V_{23} V_{31}^2}{V_{21}^2 V_{32} V_{33}} &\arg \frac{V_{21} V_{23} V_{32}^2}{V_{22}^2 V_{31} V_{33}} &\arg \frac{V_{21} V_{22} V_{33}^2}{V_{23}^2 V_{31} V_{32}} \\
\end{pmatrix}
= 
T^{2} \Phi T^{2} - T ^{2} \Phi T \, . 
\end{align}
Alternatively, by multiplying $T$ from the left, one obtains
\begin{align}
 T^{2} \D - T \D = \Phi T^{2} - \Phi T \, ,
\end{align}
and for matrix elements 
\begin{align}
 \d^{(\a+1, i)} - \d^{(\a+2, i)} = \Phi_{\a, i+2} - \Phi_{\a, i+1} \, . 
 \label{TRow}
\end{align}
Iterating these two transfers, we can find the sum rules for all combinations between $\delta^{(\a i )}$ and $\d^{(\b j)}$.

Finally, we reproduce the results of the previous work. 
Since it suffices to examine the relation between $\delta^{(13)}$ and $\delta^{(11)}$, 
the case of $\a=1, i=2$ in Eq.~(\ref{Tcolumn}) becomes  
\begin{align}
\d^{(1 1)} - \d^{(1 3)}  = \Phi_{2 2} - \Phi_{3 2} \, ,  ~~~ 
\pi -  \d_{\rm KM} - \d_{\rm PDG}  =  \a - \g \, . 
\end{align}
Thus, the relation $\delta_{\rm PDG} + \delta_{\rm KM} = \pi - \alpha + \gamma$ is confirmed. 
Furthermore, it has also been shown in the previous work \cite{Yang:2025law, Yang:2025ftl}, 
such a difference between phases and angles is expressed in terms of a third-order invariant  \cite{Luo:2025wio}
\begin{align}
 \d_{\rm PDG} - \g &= 
  \arg \left[ -  { V_{us}  V_{cd} V_{tb} \over \det V_{\rm CKM} } \right] 
= \pi - \d_{\rm KM} - \a \, .   
\end{align}
A comprehensive study of this kind of relation will be performed in the subsequent work.
The results generalize the previous sum rule to the nine CP phases in the different parameterizations 
and angles of unitarity triangles. 

\section{Summary}

In this letter, we present rephasing invariant formulae $\delta^{(\a i)} = \arg \left[ { V_{\a 1} V_{\a 2} V_{\a 3} V_{1i} V_{2i} V_{3i} / V_{\a i}^{3} \det V } \right]$ for CP phases $\delta^{(\a i)}$ associated with the nine parameterizations $V^{(\a i)}$ of a flavor mixing matrix. 
Here, $\a$ and $i$ denote the row and column with trivial phases in $V^{(\a i)}$. 
In practice, since cancellations occur between the numerator and the denominator, 
these are a necessary and sufficient set of irreducible fifth-order invariants constructed using the determinant. 

Furthermore, we find exact sum rules $ \d^{(\a , i+2)} - \d^{(\a, i+1)} = \Phi_{\a+1, i} - \Phi_{\a+2, i}$
and $ \d^{(\a+1, i)} - \d^{(\a+2, i)} = \Phi_{\a, i+2} - \Phi_{\a, i+1}$ between the nine phases $\delta^{(\a i)}$ and the nine angles $\Phi_{\a i}$ of the unitarity triangles.  
These relations revealed intrinsic underlying orders connecting the CP phases and angles, which seem to be random and independent.
We show that a relation from the previous work $\delta_{\rm PDG} + \delta_{\rm KM} = \pi - \alpha + \gamma$ is immediately reproduced as a special case ($\a=1, i=2$) of the first sum rules.  
These results provide a general and concise framework for differences between angles and phases, 
and offer a valuable guide for future experimental tests of CP violation. 

\section*{Acknowledgment}

The study is partly supported by the MEXT Leading Initiative for Excellent Young Researchers Grant Number JP2023L0013.


\end{document}